\documentclass[10pt,conference]{IEEEtran}

\newcommand{\hpcayear}{2025}

\usepackage{amsmath,amsfonts}
\usepackage{algorithmic}
\usepackage{graphicx}
\usepackage{textcomp}
\usepackage{xcolor}
\usepackage{fancyhdr}
\usepackage{comment} 
\usepackage{footnote}
\makesavenoteenv{tabular}
\makesavenoteenv{table}
\usepackage{multirow}
\usepackage{subfig}
\usepackage{hyperref}
\usepackage{makecell}
\usepackage{array}
\usepackage{tikz}

\usepackage{color,soul}
\usepackage{comment}

\usepackage{makecell}
\usepackage{multirow}
\usepackage{xspace}

\newcommand{\ballnumber}[1]{\tikz[baseline=(myanchor.base)] \node[circle,fill=.,inner sep=1pt] (myanchor) {\color{-.}\bfseries\footnotesize #1};}

\newcommand{\sys}{C4\xspace}

\title{Enhancing Large-Scale AI Training Efficiency: The C4 Solution for Real-Time Anomaly Detection and Communication Optimization}

\def\hpcacameraready

\makeatletter
\newif\if@restonecol
\makeatother
\let\algorithm\relax
\let\endalgorithm\relax
\usepackage[ruled,vlined]{algorithm2e}

\author{
  \ifdefined\hpcacameraready
    \IEEEauthorblockN{Jianbo Dong\dag, Bin Luo\dag, Jun Zhang\dag, Pengcheng Zhang\dag, Fei Feng\dag,  Yikai Zhu\dag,  Ang Liu\dag, \\ Zian Chen\dag,  Yi Shi\dag, Hairong Jiao\dag,  Gang Lu\dag,  Yu Guan\dag,  Ennan Zhai\dag,  Wencong Xiao\dag,  Hanyu Zhao\dag, \\ Man Yuan\dag,  Siran Yang\dag,  Xiang Li\dag, Jiamang Wang\dag,  Rui Men\dag,  Jianwei Zhang\dag,  Chang Zhou\dag, \\ Dennis Cai\dag,  Yuan Xie\dag\ddag,  Binzhang Fu\dag}
      \IEEEauthorblockA{
        \dag~Alibaba Group\ \ \ \ \ \ \ \ \ddag~Hong Kong University of Science and Technology\\
        jianbo.djb@alibaba-inc.com
      }
  \else
    \IEEEauthorblockN{\normalsize{HPCA \hpcayear{} Submission
      \textbf{\#\hpcasubmissionnumber{}}} \\
      \IEEEauthorblockA{
        Confidential Draft \\
        Do NOT Distribute!!
      }
    }
  \fi 
}

\fancypagestyle{camerareadyfirstpage}{%
  \fancyhead{}
  
  \fancyhead[C]{
    \ifdefined\aeopen
    \parbox[][12mm][t]{13.5cm}{\hpcayear{} IEEE International Symposium on High-Performance Computer Architecture (HPCA)}    
    \else
      \ifdefined\aereviewed
      \parbox[][12mm][t]{13.5cm}{\hpcayear{} IEEE International Symposium on High-Performance Computer Architecture (HPCA)}
      \else
      \ifdefined\aereproduced
      \parbox[][12mm][t]{13.5cm}{\hpcayear{} IEEE International Symposium on High-Performance Computer Architecture (HPCA)}
      \else
      \parbox[][0mm][t]{13.5cm}{\hpcayear{} IEEE International Symposium on High-Performance Computer Architecture (HPCA)}
    \fi 
    \fi 
    \fi 
    \ifdefined\aeopen 
      \includegraphics[width=12mm,height=12mm]{ae-badges/open-research-objects.pdf}
    \fi 
    \ifdefined\aereviewed
      \includegraphics[width=12mm,height=12mm]{ae-badges/research-objects-reviewed.pdf}
    \fi 
    \ifdefined\aereproduced
      \includegraphics[width=12mm,height=12mm]{ae-badges/results-reproduced.pdf}
    \fi
  }
  \fancyfoot[C]{}
}
\begin{document}
\maketitle

\ifdefined\hpcacameraready 
  \thispagestyle{camerareadyfirstpage}
  \pagestyle{empty}
\else
  \thispagestyle{plain}
  \pagestyle{plain}
\fi

\newcommand{\hpcaheight}{0mm}
\ifdefined\eaopen
\renewcommand{\hpcaheight}{12mm}
\fi





\begin{abstract}
The emergence of Large Language Models (LLMs) has necessitated the adoption of distributed training techniques, involving the deployment of thousands of GPUs to train a single model. 
Unfortunately, the efficiency of large-scale distributed training systems is often suboptimal due to the increased likelihood of hardware errors in high-end GPU products and the heightened risk of network traffic collisions.
Specifically, GPUs involved in the same job require periodic synchronization to exchange necessary data, such as gradients, parameters, or activations. 
As a result, any local hardware failure can disrupt training tasks, and the inability to swiftly identify faulty components leads to a significant waste of GPU resources. 
Moreover, prolonged communication due to traffic collisions can substantially increase GPU waiting times.

To address these challenges, we propose a communication-driven solution, namely the \sys.
The key insights of \sys are twofold. 
First, the load in distributed training exhibits homogeneous characteristics and is divided into iterations through periodic synchronization, therefore hardware anomalies would incur certain syndrome in collective communication. 
By leveraging this feature, \sys can rapidly identify the faulty components, swiftly isolate the anomaly, and restart the task, thereby avoiding resource wastage caused by delays in anomaly detection. 
Second, the predictable communication model of collective communication, involving a limited number of long-lived flows, allows \sys to efficiently execute traffic planning, substantially reducing bandwidth competition among these flows.
The \sys has been extensively deployed across real-world production systems in a hyperscale cloud provider, yielding a significant improvement in system efficiency, from 30\% to 45\%. This enhancement is attributed to a 30\% reduction in error-induced overhead and a 15\% reduction in communication costs.
\end{abstract}

\maketitle

\section{Introduction}

The rapid advancement of Large Language Models (LLMs)~\cite{gpt, gpt2, gpt3, llama, glm, qwen} 
has significantly impacted the fields of machine learning (ML) and 
artificial intelligence (AI). LLMs such as GPT~\cite{gpt}
and Llama~\cite{llama} families are exhibiting human-like proficiency in text generation,
question-answering, and even code creation. 
Despite the promising capabilities of these models, training them to reach their full potential is a complicated and resource-intensive process.
Training a GPT-like model that has 175 billion parameters,
for example, may require up to two months in a cluster of 1000 GPUs~\cite{traingpt}.


One of the major hurdles we faced when training LLMs is the suboptimal utilization of GPU resources, which can be attributed to two primary reasons.
First, a significant amount of time (over 30\%) is spent on recovering from system errors throughout the training lifespan. 
Extended training sessions necessitate a stable process;
however, the latest generation GPUs~\cite{a100,h100} tend to
exhibit high error rates, likely a consequence of their rapid development,
rushed delivery, and increased power consumption. 
When running in Bulk Synchronous Parallel (BSP)~\cite{bsp} mode, 
errors in any node can cause the entire job to fail.
This necessitates extensive time for system diagnosis 
and node isolation before we restart the job. 

Furthermore, tuning the training process of these models to peak efficiency is a
complex task, with communication costs being a significant impediment to
scalability.
That is, 
GPUs may experience delays at
synchronization points while awaiting the outcomes of 
collective operations.
The state of the arts~\cite{eflops,accl} suggest managing network traffic to 
improve communication performance. 
However, a shared physical cluster with multiple
tenants running concurrent training jobs adds to the complexity of these
patterns, exacerbating the difficulty in effective management.
Additionally, the appearance of downlinks, which is increasingly common in large-scales systems, 
also diminishes the efficacy of previous traffic engineering strategies.

In order to tackle these challenges, this paper introduces Calibrating Collective Communication over
Converged Ethernet (\sys), an innovative system designed to 
unleash the computational capabilities of massive collaborating
GPUs in large-scale AI clusters,
by harnessing the inherent characteristics of distributed training. 
Specifically, it minimizes  downtime caused by hardware errors by enabling the design of a highly efficient error recovery mechanism.
Leveraging the homogeneous running rhythm of the distributed workers, it enables real-time fault detection by analyzing the timing discrepancies among them at the synchronization points, namely, the collective communication operations in each step.
Moreover, \sys improves runtime communication efficiency through global traffic engineering. 
By exploiting the features of network traffic in AI clusters, \sys implements a cluster-level path management mechanism, mitigating the impact of traffic collisions and ensuring seamless collaboration among GPUs.

The contributions of this paper include:
\begin{itemize}
    \item  We extend the capabilities of Alibaba Collective Communication Library (ACCL)~\cite{accl} to support real-time status monitoring and communication path control, enabling the development of real-time error detection and global traffic engineering. 
    \item We design and implement a real-time hardware error detection subsystem that collects and analyzes the runtime status of ACCL across different workers, and builds upon this a fast recovery system that automatically isolates faulty nodes and restarts the affected jobs.
    \item We develop a global traffic engineering subsystem that manages all the networking path resources in a cluster, and intelligently controls path allocation for establishing connections in ACCL, ensuring balanced load distribution across available paths.
    \item We successfully deploy \sys in our AI clusters, serving key customers for LLM training. It effectively removes the majority of error-induced overhead, typically accounting for about 30\% of the total time. 
    Additionally, \sys mitigates the communication costs, and improves the system throughput by approximately 15\%. 
\end{itemize}

\section{Understanding the Challenges in Operational AI Clusters}
\label{sec:back}

\subsection{New Challenges}
In response to the immense success of LLMs, cloud providers have deployed dedicated AI clusters to effectively manage the growing demand for large-scale training tasks.
Unlike traditional data centers for traditional cloud applications, large-scale AI clusters face distinct challenges arising from emerging workloads and rapidly evolving hardware systems.


\emph{One of the key challenges is that distributed training jobs are highly sensitive to local failures, which occur with notable frequency in modern large-scale training systems.}
This sensitivity arises from the periodical synchronization required among workers in each training step. 
It is well-known that contemporary LLMs are capable of harnessing tens of thousands of GPUs for their distributed training processes. 
The mainstream distributed training frameworks~\cite{mega,zero} support a variety of parallel training strategies across multiple dimensions.
With these parallel training strategies, each GPU manages a fraction of the model or input for localized computations.
The workers within the same parallelism group must exchange gradients, parameters, or activations for further processing.
If an exception occurs in any worker, it will block subsequent workers and cause the entire job to crash.
Consequently, distributed training jobs, similar to HPC applications, must frequently save checkpoints and restart the entire job from the last valid checkpoint in the event of errors.
What's worse, our experience has shown that high-end GPU products tend to have higher error rates, increasing the probability  a distributed training job may crash, especially as the system scales out.
The recovery from job crashes can be costly and time-consuming, leading to a significant waste of computational capacity.

Another key challenge comes from the pronounced disparity in expected delivery standards of AI clusters. 
Specifically, users demand peak task performance, driven by two primary motivations.
First, the value of the developed models is intrinsically tied to how swiftly they can be brought to market, 
making expedited training a priority. 
Second, the significant capital 
invested in the training infrastructure means that any performance degradation translates directly into substantial financial repercussions.
\emph{However, maintaining a consistently optimal operational state for these large-scale systems is particularly challenging over extended periods, especially as the number of GPUs in a single job reaches thousands or even tens of thousands.} 

\subsection{Pain Points and Their Influences on Training Jobs}
\begin{figure}[t]
\centering
\includegraphics[width=\linewidth]{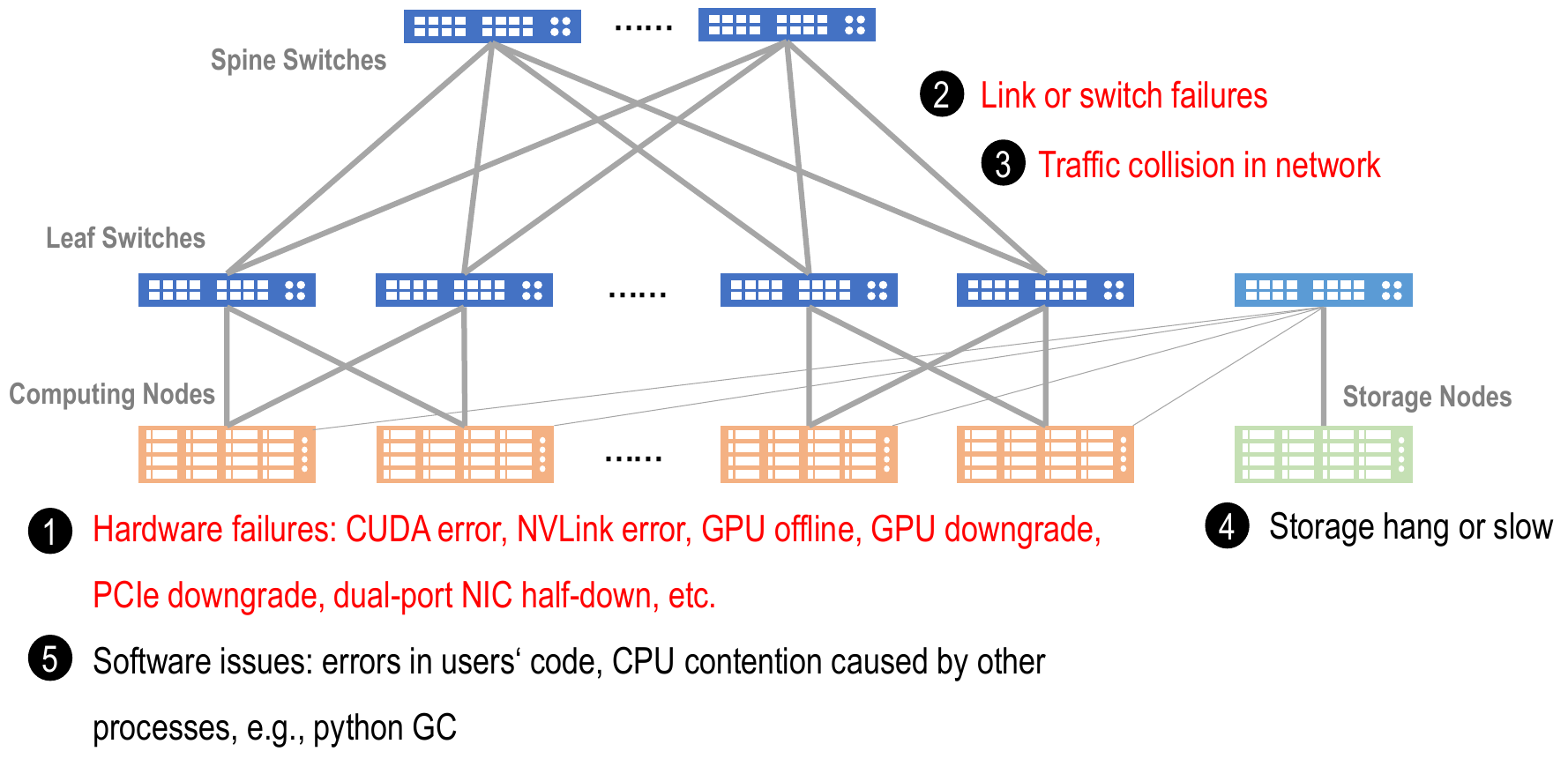}
\caption{Issues that can adversely affect system utilization.}
\vspace{-0.5cm}
\label{fig:issue}
\end{figure}

Fig.~\ref{fig:issue} illustrates the various issues that can negatively impact system utilization, including \ballnumber{1} anomalies in computing nodes, \ballnumber{2} failures and \ballnumber{3} traffic collision in networking subsystem, \ballnumber{4} I/O hang or slow in storage subsystem, and \ballnumber{5} problems in users' code or supporting software. 
First of all, nodes with critical failures may cause job crashes. And some of the computing nodes could exhibit low performance, commonly referred to as slow nodes, often due to defects in GPUs or NICs. 
The likelihood of encountering defect nodes increases with system scales, ultimately undermining the efficiency of the entire system.
Furthermore, traffic collisions among the workers, whether from the same or different jobs, can prolong the communication process and subsequently reduce system throughput. 
Link and switch failures in network subsystem could exacerbate the extent of traffic collision. 
In addition, issues in storage subsystem and software layer would systematically affect all the workers.

\begin{figure*}[t]
\centering
\includegraphics[width=\linewidth]{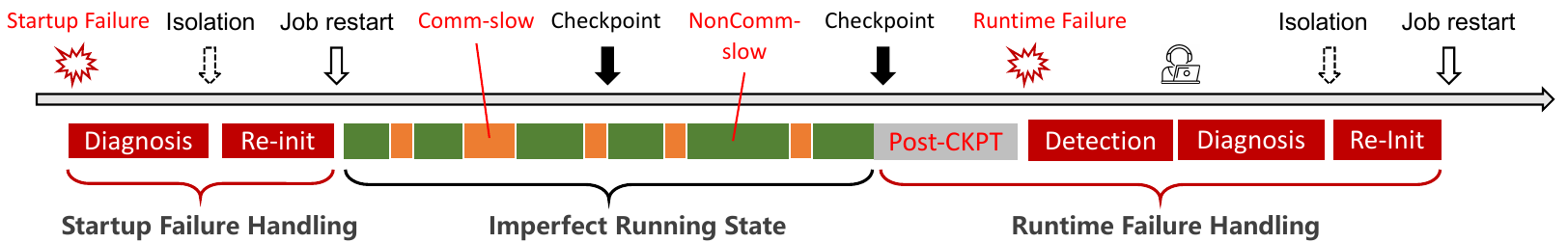}
\caption{The influences of system issues on training jobs: 1) job failures (at startup and runtime) incurred overhead, and 2) slowdowns in communication or non-communication processes (computing or data loading). }
\vspace{-0.3cm}
\label{fig:problem}
\end{figure*}

From users' perspective, system utilization is impaired in two distinct ways, as illustrated in Fig.~\ref{fig:problem}.
First, critical failures cause frequent job crashes, leading to a waste of computing power. 
Failures encountered during system startup necessitate additional time for system diagnosis, error component isolation, and job restart. 
While, if failures occur at runtime, there are additional overheads including post-checkpoint and fault detection costs.
Upon crashing, the job must revert to the last valid checkpoint, rendering the execution time after the last checkpoint useless, which we refer to as post-checkpoint cost. 
There is also a delay between the occurrence of an error and the users' awareness, known as the fault detection cost.
Besides job crashes, non-critical issues at runtime, such as performance degradation in GPUs or PCIe fabric, and network traffic collisions, can also adversely affect system throughput.
The extended communication time is referred to as communication slowdown, while the prolonged computing or data loading time is referred to as non-communication slowdown.
Any overhead in communication or non-communication indicates degradation in system throughput.
We would discuss the causes and effects of job crashes and runtime slowdowns respectively.

\subsection{The Quantitative Analyses of Job Crashes.}

To investigate the underlying causes of job crashes and how the system utilization is impacted, we chose a representative job (one of jobs with largest scale) for analysis.
This particular job, running in a newly deployed cluster, witnessed a total of 40 crashes within one month. 
We categories these crash causes and show their distribution in Tab.~\ref{tab:errors}.

\emph{The table reveals that it's challenging to discern the underlying causes from the user's perspective, as the majority of them simply show themselves as ``NCCL Errors''.} This is largely due to the synchronous nature of distributed training tasks. For instance, a hardware issue in a GPU could lead to the crash of one of the worker processes, causing its peer workers to fail to receive an acknowledgment after sending a message to the crashed worker. Consequently, this triggers an error code of 12 in the communication library. This same error code could also be generated by other issues such as network disconnections, memory overflows resulting in container eviction, or users' code errors like tensor size mismatches. 
Regrettably, the industry lacks efficient tools to identify the root causes of these errors in a timely manner.

\begin{table}[t]
\caption{The distribution of crash causes recorded over a one-month from a representative job employing 4,096 GPUs.}
\label{tab:errors}
\centering
\footnotesize
\begin{tabular}{lllll}
\hline
\textbf{Users' View} & \textbf{Root Causes} & \textbf{Proportion} & \textbf{Local}\\ \hline \hline
NCCL Error & CUDA Error        & 12.5\%  & 100\% \\ \hline
NCCL Error & ECC/NVLink Error  & 27.5\%  & 100\% \\ \hline
NCCL Error & NCCL timeout      & 20\%    & 75\% \\  \hline
NCCL Error & ACK timeout       & 27.5\%  & 81.8\%\\ \hline
Network Error & Others         & 12.5\%  & 40\%\\   \hline
\vspace{-0.7cm}
\end{tabular}
\end{table}

\emph{This table also reveals an important fact: the majority of errors (approximately 82.5\%) are confined to specific nodes or individual devices, largely attributed to the elevated error rates in high-end GPU productions.}
This indicates the potential for isolating the localized faulty components, presenting an opportunity to tackle the majority of errors\footnote{Other systematic faults, such as faults in distributed storage subsystem, would be handled by dedicated monitoring tools}. 
Nevertheless, due to the similarity in symptoms for most failures, manually identifying the root causes and locating the defective nodes could consume hours to days, particularly in a cluster with thousands of GPUs.

\begin{figure}[t]
\centering
\includegraphics[width=\linewidth]{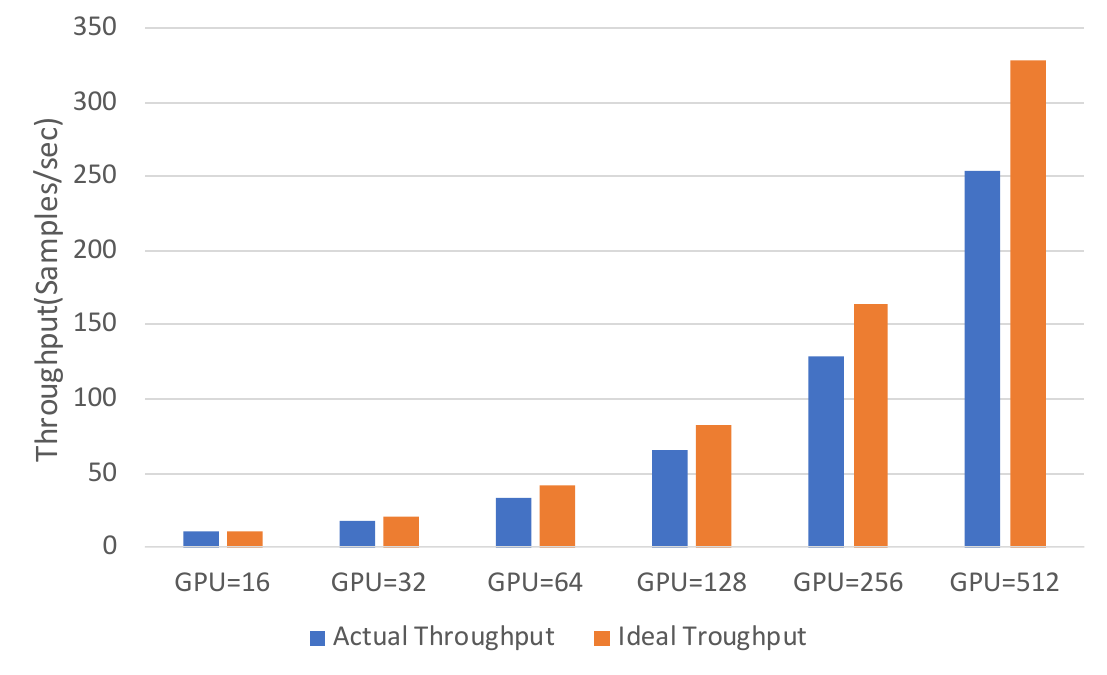}
\caption{Performance loss grows with system scale, due to the increased extent of traffic collision.}
\vspace{-0.6cm}
\label{fig:perfproblem}
\end{figure}

In our experience, around 30\% of our time was dedicated to error detection, system diagnosis, isolating defective nodes, and restarting processes, leaving only 70\% for actual computing tasks, in the absence of efficient fault detection and diagnostic tools. 
To minimize the failure-incurred costs, it is essential to address each contributing factor appropriately. 
For swift diagnosis, our system automatically identifies defective nodes by leveraging the 
inherent characteristics of training jobs, such as balanced computing costs 
within a data parallel group and synchronization between iterations. 
Upon locating a defect, 
the system promptly initiates isolation and restart procedures, 
deferring in-depth root cause analysis to offline processing.
By developing automatic fault detection tool, we can eliminated human intervention, 
expediting the process from tens of minutes to mere seconds. 
In addition, to curtail the post-checkpoint cost, 
we have implemented a rapid 
checkpoint mechanism, similar to the prior work~\cite{gemini}, 
capable of saving checkpoints approximately every 10 iterations.

\subsection{The Quantitative Analyses of Runtime Slowdowns.}
\label{sec:runtime}

Since non-critical failures in individual nodes can be addressed using the same strategy as critical errors, we will put emphasis on communication slowdowns in network.
To begin with, we designed a specialized high-performance network architecture for AI clusters, with each computing node equipped with multiple high-performance Network Interface Cards (NICs) supporting Remote Direct Memory Access (RDMA) to facilitate fast inter-node communication among the heavily packed GPUs in each node. 
Each NIC provides two physical ports, connected to a pair of leaf switches, which are interconnected to spine switches through the Fat-Tree topology, with a 1:1 oversubscription ratio. 
This architecture offers distinct advantages:
It provides high system availability and maintainability, allowing users to react more flexibly to non-critical link failures, 
such as scheduling timely restarts or initiating on-demand checkpoints that are strategically based on recent checkpointing activities. 
Additionally, it expands the maximum system scale by doubling the number of spine switches in the system, consequently doubling the total computational capacity of GPUs.

Just as a coin has two sides, this architecture also presents challenges in communication performance.
\emph{It's known that elevated communication costs may arise from traffic collision among various flows.}
When numerous flows contend for the limited bandwidth of a single link, the flow completion time can increase by several times.
In collective communication, any flow that is throttled can have a ripple effect, hindering the entire communication group from proceeding with subsequent operations.
\emph{Moreover, a new problem arises with dual-port NICs. 
If there is an imbalance in the loads on the two physical ports, the port with the higher load can become a bottleneck.}
For today's distributed training jobs, these issues can significantly prolong the communication latency and lead to degradation of communication performance.
As modern models and the training clusters grow in scale, this problem would be further exacerbated.
Fig.~\ref{fig:perfproblem} presents a comparison of actual versus ideal performance,
training a representative GPT-like model composed of 22 billion parameters.
It is evident that there exists a discrepancy between the effective performance achieved and the theoretical ideal performance, 
and this disparity amplifies as the scale of the system increases. 
Notably, when the system scales to 512 GPUs, the effective performance drops to 30\% below the ideal performance.

\emph{Our analysis reveals that communication patterns in AI clusters markedly differ from those in traditional cloud environments. }
In conventional cloud settings, we commonly observe a large number of 
concurrent connections, often several tens of thousands per instance. 
Such a volume of connections can be distributed 
across the network with relatively low variation using the Equal-Cost Multi-Path (ECMP)~\cite{ecmp} routing strategy.
Conversely, in distributed training systems, there are only a few longer-lasting connections, with each node typically managing around a few hundred connections. 
Traffic collisions in this environment can lead to substantial variations in the effective bandwidth of these connections, resulting in multi-fold increases in communication delays.
However, this challenge is accompanied by an opportunity: the limited number of connections and their predictable patterns in AI clusters offer a unique advantage for enhancing communication efficiency through comprehensive global traffic engineering.

In a nutshell, the effective performance of large-scale training clusters
is significantly influenced by hardware defects and traffic collisions. 
To maximize GPU utilization, it is crucial to reduce the time spent on fault detection and system diagnosis, thereby minimizing downtime. 
Additionally, eliminating unnecessary communication overhead is vital to prevent GPU stalls during periodic synchronization, ensuring optimal runtime performance.

\section{Design}
\label{sec:design}

To tackle the key issues identified earlier, i.e., the overheads induced by local failures and unnecessary slowdowns in communication and other processes.  
We will explain the rationale behind our choice of system architecture, the compromises we have accepted, and offer an in-depth analysis of the particular technical hurdles we encountered.

\subsection{Mitigating Error-induced Downtime}
\label{sec:design1}

As demonstrated in the previous section, the significant hardware instability in computing nodes is a primary factor contributing to the inefficiency of AI clusters. 
To mitigate the impact of hardware failures, we have adopted a two-fold approach: 1) reducing hardware failure rates and 2) systematically tolerating such failures. 
\emph{While the first aspect is beyond the scope of this paper, we briefly mention the importance of temperature control. }
Techniques such as Dynamic Voltage and Frequency Scaling (DVFS) have been utilized to prevent GPU overheating, although they can impact performance consistency. 
We have enhanced cooling through increased fan speed and improved air conditioning, allowing for temperature regulation with minimal cost impact. 
The industry is also moving towards more efficient cooling solutions such as liquid cooling for future AI infrastructures.

With respect to the second aspect, we have collaborated with relevant teams to explore multiple solutions. 
Traditional cloud applications typically utilize online fault tolerance techniques. 
These may involve withstanding computational faults through redundant computations~\cite{lockstep, redundant, reunion}, ensuring reliable storage with erasure coding (EC)~\cite{erasure, erasure2,erasure3,erasure4} and/or triple-replica techniques, and mitigating network anomalies with multi-path (dual uplinks) strategies~\cite{mptcp,mprdma}. 
For high-performance computing applications, the prevailing approach involves offline fault tolerance methods~\cite{checkpoint}. 
This includes the periodic saving of checkpoints, allowing tasks to be restarted from the last checkpoint in the event of system failure.
Yet, any computations made between the last checkpoint and the time of the failure will be lost. 
In our large-scale parallel training systems, we have implemented a practical and hybrid technical strategy. 
This includes adopting proven online fault-tolerant technologies in distributed storage and networking systems. 
While for computing resources, it's more cost-effective to use backup resources and offline fault-tolerant techniques, given the pricing of GPU products. 
As a result, we have allocated 64 backup GPUs across 8 servers for every 1024 GPUs on 128 servers, ensuring consistent communication and performance for parallel training on any of the 128 servers within this 136-server pool.

\begin{figure}[htbp]
\centering
\includegraphics[width=\linewidth]{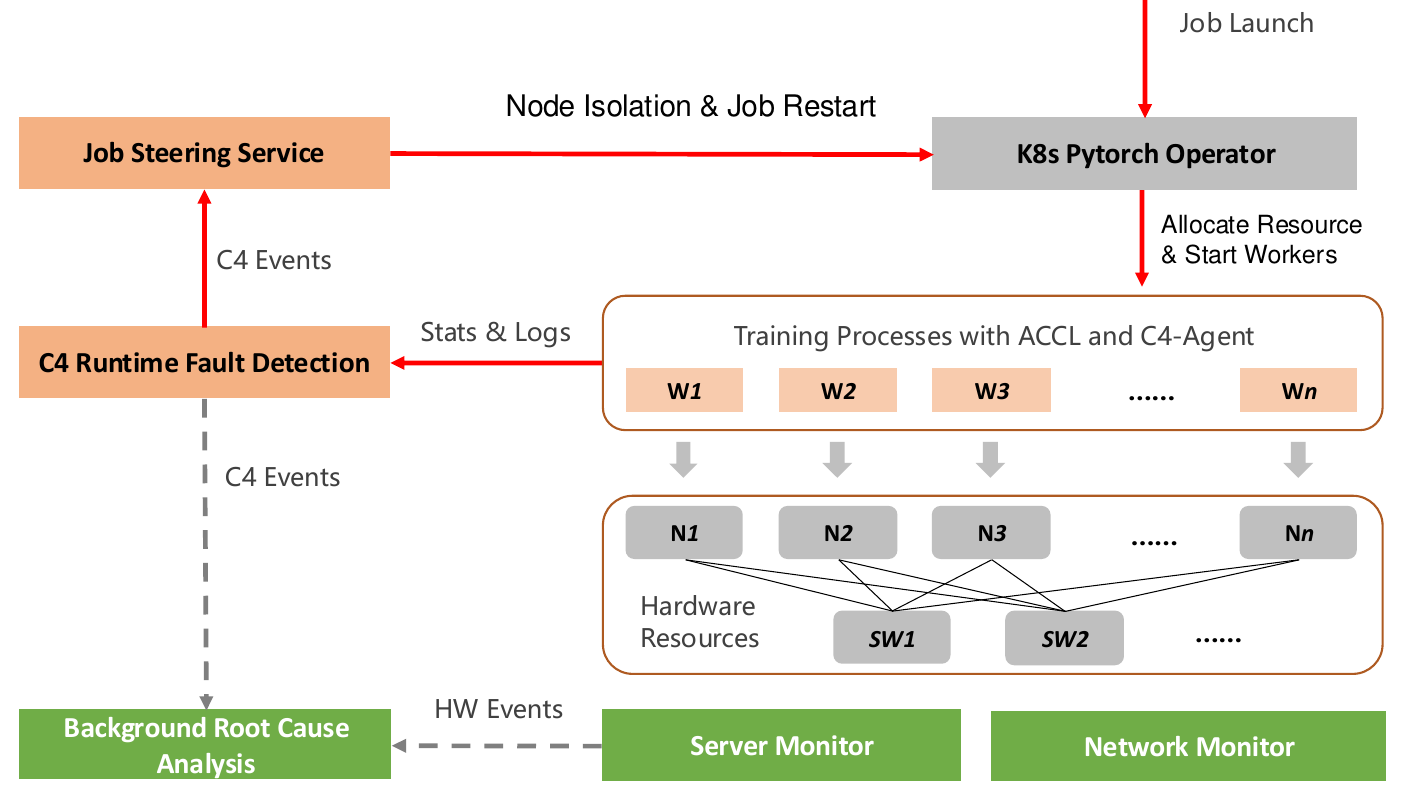}
\caption{The working flow of C4D.}
\label{fig:diag}
\end{figure}

\textbf{Our approach: C4D.} 
The system that we designed to handle computational faults is shown in Fig.~\ref{fig:diag}. The core of this system is the C4D (C4 Diagnose) subsystem, designed to quickly detect hardware failures for prompt task restarts.
Initially, C4 agents monitor the operational status of training workers and transmit the data to a centralized master. The master then evaluates the well-being of the training workers by comparing the gathered information. If any irregularities are detected, it informs the job steering service to isolate the problematic nodes and restart the job from the most recent valid checkpoint. Simultaneously, it also sends the events to a background root cause analysis system for offline diagnosis.

\begin{figure}[h]
\centering
\includegraphics[width=\linewidth]{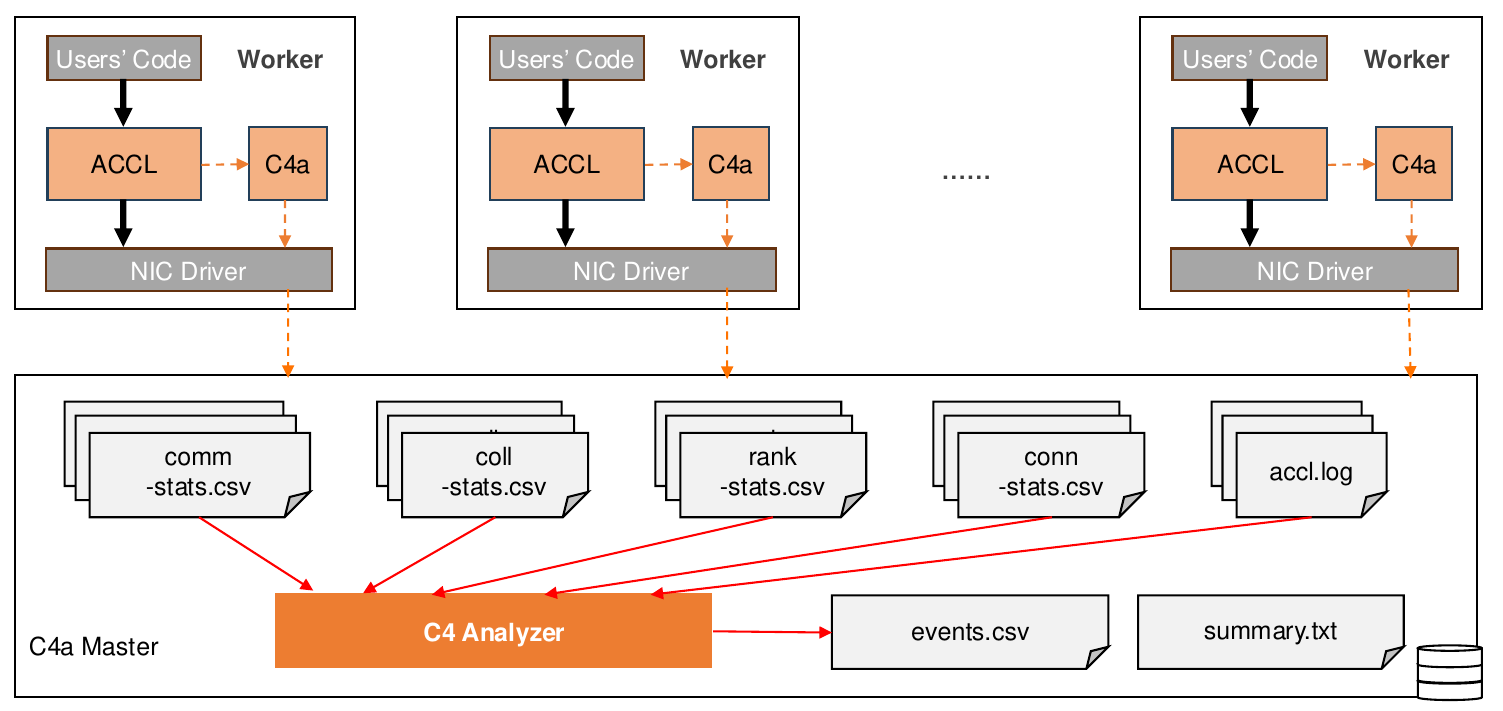}
\caption{The architecture of C4D.}
\vspace{-0.5cm}
\label{fig:c4d}
\end{figure}

The effectiveness of C4D is grounded in two key insights: (1) parallel training tasks exhibit a regular and predictable running rhythm, enabling us to precisely identify abnormal behaviors; (2) the Bulk Synchronous Parallel (BSP) execution model in parallel training requires regular synchronization points. And these synchronization points are used as anchors for measuring anomalies.
Building on these insights, we developed the C4D to facilitate online fault detection, as shown in Fig.~\ref{fig:c4d}.
The key elements of the C4D includes the enhanced communication library ACCL, a central C4D master, and the C4a (C4 agent) that acts as an intermediary.
Here, ACCL is enhanced to provide the capability for online monitoring of collective communication operations.

\begin{figure}[h]
\centering
\includegraphics[width=\linewidth]{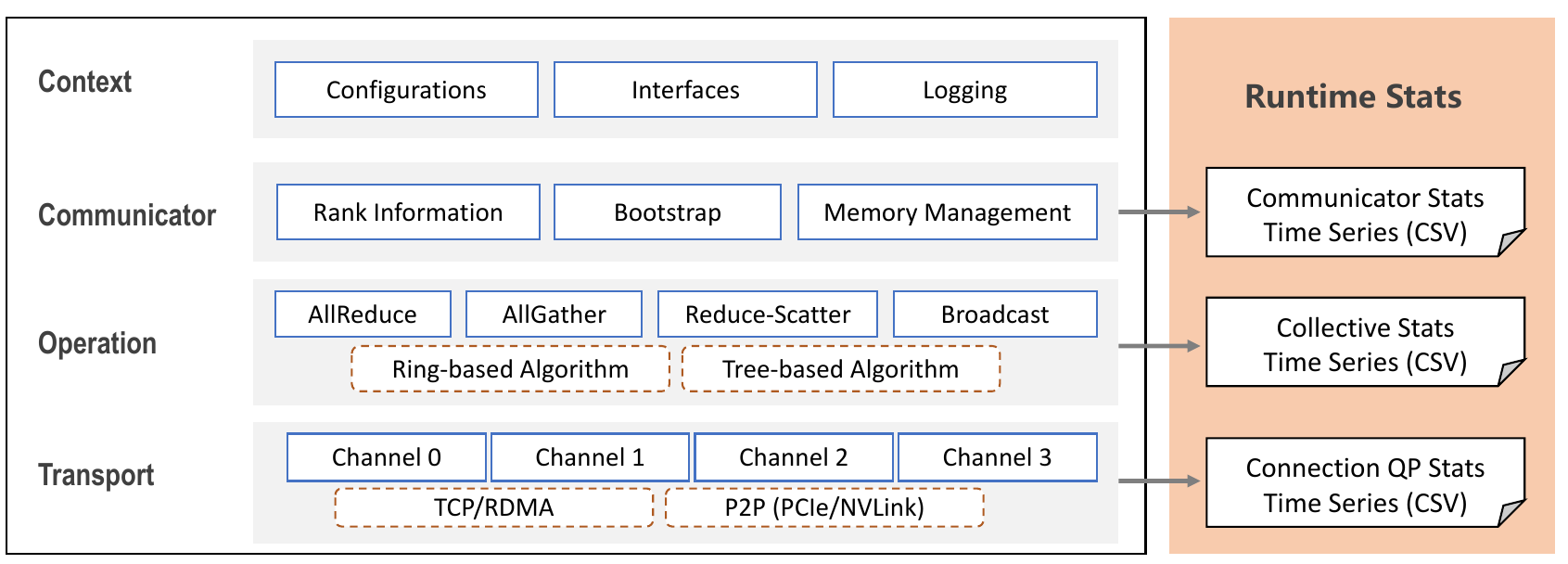}
\caption{The Enhancement of ACCL.}
\label{fig:eccl}
\end{figure}

\textbf{C4D Monitoring.} Fig.~\ref{fig:eccl} shows the architecture and enhancement in ACCL. 
We have extended the monitoring capabilities of the bottom three layers, namely the communicator, operation, and transport layers, in a top-down fashion. 
We gather information on communicators, encompassing communicator IDs, involved devices, and device ranks. 
Additionally, we record details about collective operations, such as operation types, communication algorithms, data types, element counts, and duration. 
Furthermore, we track the startup and completion of specific collective operations and assign each operation a sequence. 
Lastly, we collect data in the transport layer, including connection information (e.g., source and destination IPs, QP numbers, and source ports), as well as message details like message counts, sizes, and transfer duration.
Notably, it's not trivial to collect above information at runtime, with low cost and high accuracy. To precisely monitor communication kernel execution patterns, we refined all related CDUA kernels to log their start and completion times directly, as CPU timestamps and CUDA events are ineffective or inaccurate for this purpose.

\textbf{C4D analysis.} By leveraging the gathered information from ACCL, we can build the automatic error-detection system to deal with the four common types of errors that frequently occur in our clusters, i.e., \emph{communication hang}, \emph{non-communication hang}, \emph{communication slow}, and \emph{non-communication slow}.
Detecting the first two error types is relatively easy and will not be discussed in-depth here, and we will focus on the more complex task of identifying instances of the \emph{slow} syndrome. 

\begin{figure}[h]
\centering
\includegraphics[width=\linewidth]{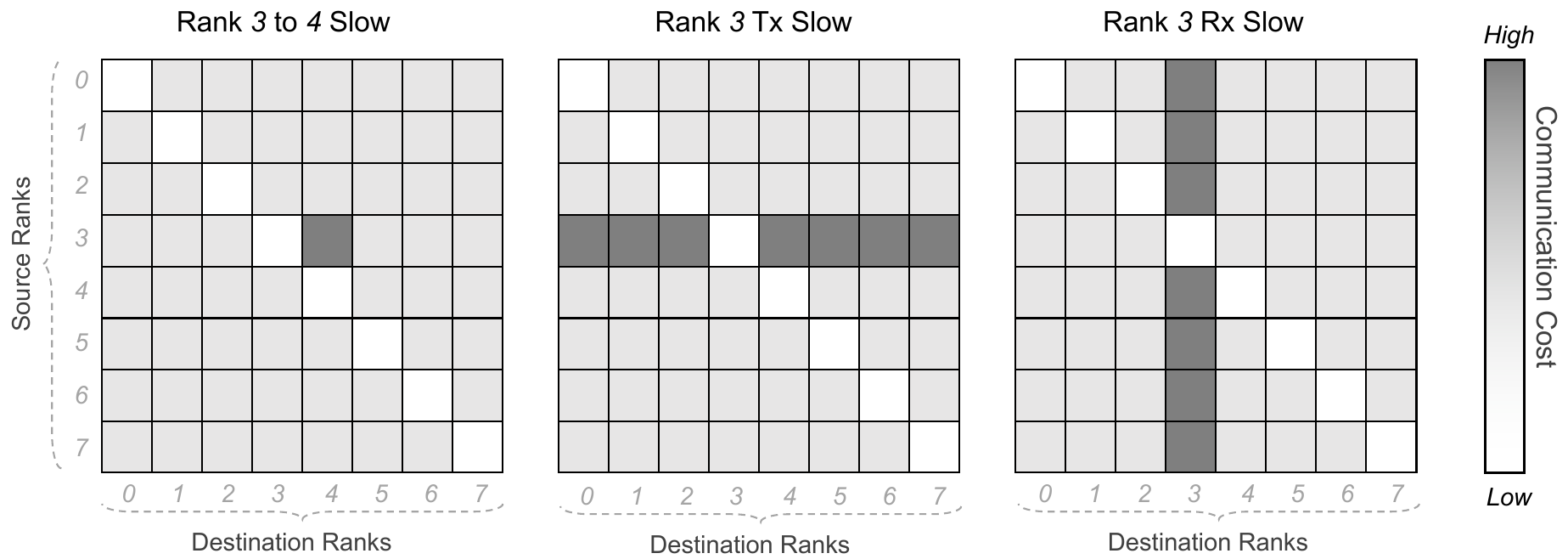}
\caption{The symptom for communication slow.}
\label{fig:commslow}
\end{figure}

\emph{Communication Slow Detection}: 
Before delving into the details, we will give a brief introduction to the data flow in collective communication. 
Considering the \emph{allreduce} operation in data parallel phase as an example, where workers need to average gradients across all model replicas. 
Frameworks and CCLs will split messages into chunks and submitting them to the network sequentially. 
Since all workers adhere to the same data splitting rules, ensuring that message sizes remain consistent when posted to the network.
Therefore, we can identify communication slowdowns at the transport layer by monitoring and comparing the completion time of messages in different workers.
An illustrative example is shown in Fig.~\ref{fig:commslow}. 
Communication delays across workers are mapped into a two-dimensional matrix, where each element represents the delay between a pair of workers, identified by their y-coordinate (source) and x-coordinate (destination). 
Large numbers in the matrix help identify slow connections: a single large number indicates a specific connection bottleneck, a row of large numbers suggests a problem with the source, and a column of large numbers points to an issue with the destination.

\emph{Non-Communication Slow Detection}:
Consider the \emph{ring-based} algorithm in \emph{allreduce} operation as an example.
In this configuration, all participating workers are connected sequentially, forming a ring-like structure. 
Each worker only communicate with its immediate neighbors in the sequence, specifically the ``previous rank'' and the ``next rank''.
Particularly, workers receive a data chunk from the ``previous rank'', perform a reduction operation with their local data chunk, 
and then pass the resultant data to the ``next rank''. 
In fact, due to the requirement that the receiver must first prepare the receive buffer and notify the sender before data transmission can occur, there is an implicit "receiver-driven" scheduling logic. 
Hence, successful communication completion in each step depends on both the sender and receiver being in a ready state.
This implicit dependency is then spread to following receivers, forming a chain of pending receivers. 
By comparing the wait time of receivers, we can pinpoint the ranks that are experiencing non-communication slows, such as those caused by additional computation or data loading costs.

\subsection{Mitigating Communication Cost}
\label{sec:design2}
%
%
%

Parallel training performance depends on the efficiency of single-node computation, data access, and collective communication. While single-node efficiency can be improved through optimizations like mixed precision~\cite{mixprec} and transformer engine~\cite{transengine}, and data access efficiency through caching mechanisms like Alluxio~\cite{alluxio} , this paper focuses on collective communication efficiency, a key factor in training scalability.

If we consider network bandwidth as a resource, optimizing collective communication performance is equivalent to finding an optimal method for resource allocation. 
In fact, collective communication can be seen as a collection of one-to-one communications between two of the workers, which may also involve computation if a reduce operation is included. 
Therefore, the problem of finding an optimal resource allocation can be broken down into two issues. 
First, we need to minimize the resource requirements of each one-to-one communication. 
Second, we need to map each one-to-one communication to network resources in a way that minimizes the total communication time.

The first issue is not the focus of this paper. But for the sake of completeness, we provide a brief introduction here. 
Once the data volume for one-to-one communication is fixed, its consumption of network resources is directly proportional to the distance it travels through the network.
To reduce the data transmission distance, we employ two optimization strategies.
First, we minimize the network diameter by leveraging high-speed NVLink interconnects, which are inherently part of the network. 
Second, we utilize topology-aware scheduling techniques to ensure that the two ranks needing to communicate are as close as possible within the network.
These optimizations mitigate the probabilities of traffic collision, and are effective for small-scale jobs.

Furthermore, we have also integrated intra-job traffic engineering techniques~\cite{eflops,accl} to prevent traffic collisions. 
However, in the case of larger AI clusters with multiple concurrent jobs, there is a lack of coordination among independent jobs. 
Consequently, additional optimization is required to deal with inter-job traffic collision.

\begin{figure}[h]
\centering
\includegraphics[width=\linewidth]{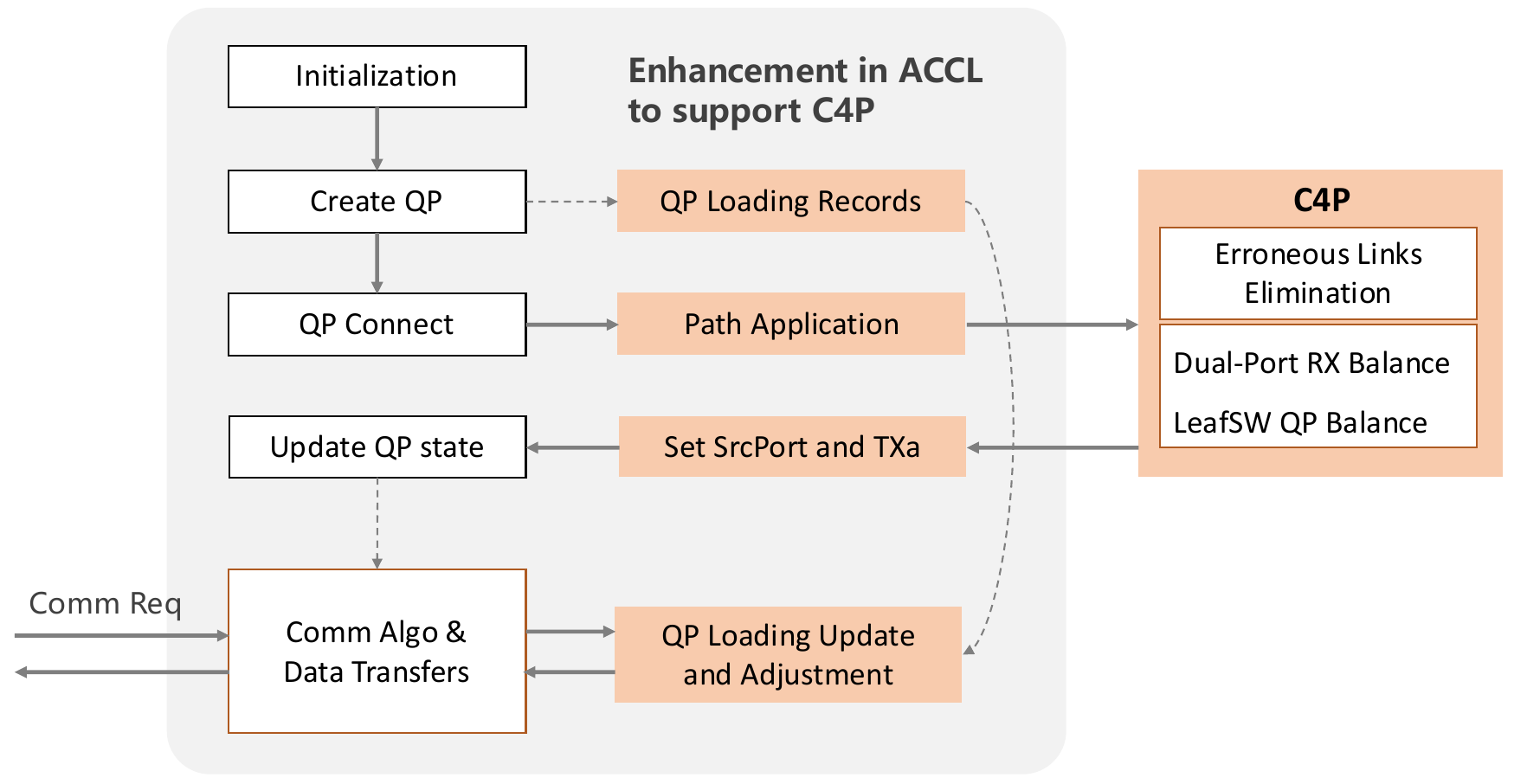}
\caption{The architecture of C4P.}
\label{fig:c4p}
\end{figure}

\textbf{Our approach: C4P.} To address this problem, we introduce the C4P (C4 Performance) system, engineered to reduce unnecessary communication costs. 
Essentially, C4P is a cluster-scale traffic engineering technique that aims to minimize network congestion by regulating the transmission paths of each data flow within the network. This concept isn't new, but it's not commonly employed in traditional data centers, because of the massive connections in network. C4P is practical in parallel training scenarios mainly because the traffic characteristics generated by parallel training are significantly different from those of conventional cloud applications. That is, parallel training tasks involve a small number of data flows but transmit large volumes of data. This presence of a few elephant flows makes it feasible for C4P to meticulously plan the route for each data flow.


The software architecture of C4P follows that of C4D but with key differences, as shown in Fig.~\ref{fig:c4p}. 
Firstly, the C4P master acts as a control center for multiple jobs or tenants, unlike the single-job-focused C4D master. 
Additionally, the C4P master records the numbers of allocated connections on each path, and allocates path for new connections considering the occupied network resources.
At last, the CCL is enhanced to support issuing path allocation requests for communicating workers and set the source port accordingly. 
\emph{The key functionalities of C4P can be summarized as: (1) identifying and avoiding faulty links at task start-up, (2) balancing RDMA QPs, i.e., connections, across healthy paths for load distribution, and (3) dynamically adapting QP workloads in response to network changes and traffic conflicts.}

The detailed working flow is as follows:
In line with previous art~\cite{accl}, we utilizes path-probing to support global traffic engineering. 
Using this method, we can identify the source ports that will direct traffic along specific paths and verify the integrity of those paths.
Initially, C4P identifies and eliminates faulty links between leaf and spine switches to establish a sub-network composed of healthy links. 
To do this, the C4P master performs full-mesh path probing via randomly selected servers per leaf switch, identifying and cataloging the healthy paths.
Upon establishing connections, ACCL submits path allocation requests to the C4P master, which replies with the source ports of RDMA connections.
The master ensures traffic from the same NIC is balanced between left and right ports by forbidding the paths from left ports to right, and vice versa.
Additionally, traffic from servers under the same leaf switch is distributed over all available spine switches to prevent network congestion. 
The ACCL constantly evaluates message completion times on various paths and prioritizes the fastest for data transfer. 
This approach helps maintain low transmission delays despite path errors or traffic congestion.

\section{Evaluations}

\subsection{Setup}
As shown in Table~\ref{tab:conf}, we outline the system configurations of our operational clusters, 
from which we have gathered the evaluation data for C4D. 
The assessments pertaining to C4P, on the other hand, were carried out on a subset of this cluster.
To prevent any interference with other ongoing jobs in the cluster and to ensure the integrity of our evaluation results, 
we have allocated a specific segment of the cluster as a controlled testing environment. 
This testbed comprises 16 nodes, equipped with a total of 128 GPUs. 
And all the nodes are directly connected to 8 dedicated leaf switches, 
ensuring an exclusive environment for our testing activities.

The nodes within this high-performance cluster are outfitted with 8 NVIDIA H800 GPUs and 8 BlueField-3 NICs~\cite{bluefield}. 
Each NIC provides two physical 200Gbps ports, which are bonded to create a single logical 400Gbps port. 
These NICs are integrated into a 3-Tier Clos network~\cite{clos}, configured in a Fat-Tree topology with a 1:1 oversubscription rate.
Leveraging Broadcom's Trident4~\cite{td4} as leaf and Tomahawk4 as spine switches~\cite{th4}, 
the cluster is capable of supporting over 10,000 GPUs. 
In a single pod, which constitutes a two-tier subnet, it can accommodate 512 GPUs.

\begin{table}[t]
\caption{Configurations}
\label{tab:conf}
\centering
\footnotesize
\begin{tabular}{lll}
\hline
Benchmarks for C4D & GPT model, 175 billion parameters  \\ \hline 
\multirow{4}{*}{Benchmarks for C4P} & allreduce operations  \\ 
& GPT model, 22 billion parameters \\
& Llama model, 13 billion parameters \\
& GPT model, 175 billion parameters\\
\hline 
Frameworks & Megatron-LM~\cite{mega}, DeepSpeed~\cite{zero}  \\ \hline 
Nodes & NV H800 * 8, BlueField3 * 8 (200Gbps * 2) 	      \\ \hline
Network & 3-Tier Clos, Fat-Tree, 1:1 oversubscription rate    \\ \hline
\end{tabular}
\end{table}

The effectiveness of C4D is evaluated with one of our real-world LLM training workloads. 
And the model for this job contains 175 billion parameters.
Training with 2400 GPUs, it requires more than a month for a complete training cycle from scratch.
In contrast, the efficiency of C4P is assessed using both collective communication benchmarks and three representative LLM training jobs. 
To ensure unbiased evaluation results for the communication benchmarks, we configured them to utilize a ring-based algorithm. 
The specific configurations of the models involved in the testing, such as parallel strategy and zero-optimizer level, will be detailed alongside the presentation of the evaluation outcomes.

\subsection{Results}

\subsubsection{The Evaluations on C4D's Effectiveness.}

We have reviewed the historical logs of the jobs from one of our inner customers, 
including their start and terminated timestamps, and whether a job is succeeded or failed by some reasons.
Our primary emphasis has been placed on a representative job that demands 2400 GPUs and necessitates over a month to complete.
As shown in Table~\ref{tab:downtime}, we compare the error-induced downtime of this job before (Jun, 2023) and after (Dec, 2023) the proposed fault-tolerant system was deployed.
The data clearly indicates that the deployment of C4D led to a substantial decrease in downtime, with an approximately 30x reduction, from 31.19\% down to just 1.16\%.

\begin{table}[t]
\caption{Error-induced Downtime Statistics}
\label{tab:downtime}
\centering
\begin{tabular}{|l|l|l|l|}
\hline
\multicolumn{4}{|c|}{\textbf{Error-induced Downtime in Jun, 2023}}\\ 
\hline
\hline
Post-Checkpoint               & 7.53\%                        & \multicolumn{2}{l|}{ } \\ 
\cline{1-2}
Detection                             &  3.41\%                        & \multicolumn{2}{l|}{ } \\ 
\hline
\multirow{5}{*}{\makecell[l]{Diagnosis \& \\ Isolation}}   & \multirow{5}{*}{19.65\%}   & ECC/NVLink Error & 8.34\%  \\
\cline{3-4}
& & CUDA Error & 4.19\%                                           \\     
\cline{3-4}      
& & CCL Timeout & 3\%                                           \\  
\cline{3-4}       
& & ACK Timeout & 1.8\%                                          \\
\cline{3-4}
& & Unknown & 2.29\%                                          \\ \hline
Re-Initialization                & 0.6\%                         & \multicolumn{2}{l|}{ } \\         
\cline{1-2}      
Total                               & 31.19\%                         & \multicolumn{2}{l|}{ } \\ 
\cline{1-4}

\multicolumn{4}{c}{ } \\

\hline
\multicolumn{4}{|c|}{\textbf{Error-induced Downtime in Dec, 2023}}\\ 
\hline
\hline
Post-Checkpoint               & 0.23\%                        & \multicolumn{2}{l|}{ } \\ 
\cline{1-2}
Detection                             &  0.05\%                        & \multicolumn{2}{l|}{ } \\ 
\hline
\multirow{5}{*}{\makecell[l]{Diagnosis \& \\ Isolation}}   & \multirow{5}{*}{0.73\%}   & ECC/NVLink Error & 0.2\%  \\
\cline{3-4}
& & CUDA Error & 0.1\%                                           \\     
\cline{3-4}      
& & CCL Timeout & 0.23\%                                          \\
\cline{3-4}
& & ACK Timeout & 0.1\%                                           \\  
\cline{3-4}       
& & Unknown & 0.1\%                                          \\ \hline
Re-Initialization                & 0.15\%                         & \multicolumn{2}{l|}{ } \\         
\cline{1-2}      
Total                               & 1.16\%                         & \multicolumn{2}{l|}{ } \\ 
\cline{1-4}

\end{tabular}
\end{table}

The downtime in in June 2023 is broken down into four parts, according to the classification in Fig.~\ref{fig:problem}.
It reveals that the majority of the downtime was due to extended periods of system diagnosis.
This was largely because identifying the precise cause of job failure could take hours or even days in the absence of our diagnostic tools.
The second major contributor to downtime was post-checkpoint overhead.
That is because the users often scheduled infrequent checkpoints to save the model parameters, not anticipating high error rates in a large-scale training cluster. 
This led to a significant amount of computing activity being rendered useless between the last valid checkpoint and the moment the job failed due to an error.
To address this issue and reduce the amount of wasted computing time, 
users have begun saving checkpoints more frequently by taking advantage of the high-performance infrastructure provided~\cite{gemini}. 
Besides the diagnosing and post-checkpoint cost, there is also 3.4\% of time spent on crash detection.
Typically, users are not immediately aware that their job has stalled following an error event.
For instance, PyTorch jobs that might experience a hang for up to 30 minutes, until the elastic agent~\cite{ea} kills the processes.
This delay creates a span of idle time for computing resources, and thus constitutes a non-negligible portion of the total downtime.

In December 2023, despite system diagnosis continuing to account for the majority of downtime, 
its contribution had been reduced by approximately 27x compared to past performance. which is somewhat lower than the average efficiency gains observed. 
The deployment of C4D has markedly accelerated both error detection and the pinpointing of faulty components, 
cutting down the response time to mere tens of seconds. 
Nevertheless, additional minutes are still required by the steering service to isolate the affected nodes and restart the job, indicating room for further improvement.
The second leading cause of downtime remains the post-checkpoint duration. 
However, with the adoption of more frequent checkpointing, users can now save checkpoints every 10 minutes. 
As a result, the post-checkpoint downtime has been significantly reduced by 33x.
The cost associated with job re-initialization has remained constant. 
It has decreased from 0.6\% to 0.15\% of the total time, 
because the average error rate has decreased by 3.33x, 
after the most vulnerable components of the system were identified and enhanced during this six months. 

Upon delving deeper into the diagnosis and isolation overhead, it has become apparent that the majority of errors are attributable to GPU defects.
These include GPU ECC Errors, NVLink Errors, and CUDA Errors.
In June 2023, such GPU-related issues were responsible for 12.53\% of the total downtime, which constituted around 2/3 of the overall overhead.
Notably, by December 2023, the incidence of these GPU-related errors had been reduced by 3.2x, 
while the associated time overhead saw a more dramatic decrease by 41.8x. 
This significant improvement can be credited to the effectiveness of C4D in handling these specific types of errors, 
allowing for a substantial reduction in corresponding overheads.
Regarding other types of errors that are only partially manageable by C4D, 
there was still a noteworthy improvement. 
The frequency of these errors was decreased by 3.4x, and the time spent addressing them saw a reduction by 16.5x. 
These numbers underscore the positive impact of C4D on overall system performance and reliability.

\subsubsection{The Evaluations on C4P's Effectiveness.}

\begin{figure}[t]
\centering
\includegraphics[width=0.75\linewidth]{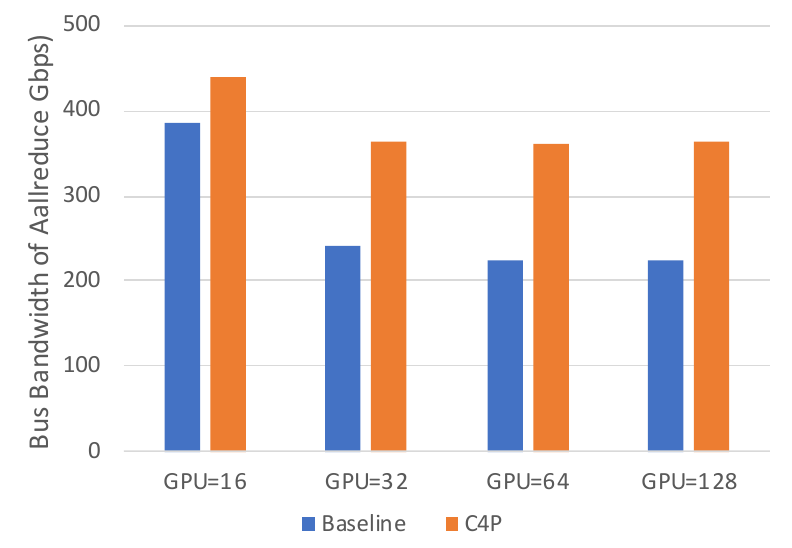}
\caption{Improvement from balanced traffic between bonded ports.}
\vspace{-0.5cm}
\label{fig:rxbalance}
\end{figure}

\textbf{Balancing traffic between the two bonded physical ports.}
As discussed in Section~\ref{sec:runtime}, when two flows are dispatched from two distinct physical ports of a NIC, 
there is a possibility that both flows could be routed to the same physical port on the receiving end, 
leading to an imbalance of traffic between the two physical ports in receivers. 
By employing C4P, we can designate a dedicated path for the flows in each physical port, thus preventing performance degradation caused by such imbalances.
Running ``nccltest'', the collective operation benchmarking tool supplied by NVIDIA, 
we are able to measure the average \emph{busbw}, a metric that reflects the effective communication performance (higher \emph{busbw} means lower communication delay).
Fig.~\ref{fig:rxbalance} shows the performance of a single \emph{allreduce} operation with and without C4P enabled.
Here, the results from the two-node (16-GPU) test case don't align with the effective bandwidth observed in the network. 
This discrepancy can be attributed to the methodology employed to calculate \emph{busbw} within the benchmarks.
We will not delve into this particular issue in this paper, and instead concentrate on other test cases to better illustrate our research outcomes.
Without C4P, the effective \emph{busbw} is lower than 240 Gbps in most test cases, which is far from the ideal bandwidth of network (around 360 Gbps).
However, with C4P activated, the effective \emph{busbw} increases close to the peak value 360 Gbps, 
which means 50\% performance gain,
showcasing the tangible benefits of C4P in optimizing network performance.

\textbf{Balancing traffic among multiple jobs.}
To demonstrate the efficacy of C4P in alleviating traffic collisions across multiple concurrent jobs, 
we conducted an evaluation involving 8 simultaneous \emph{allreduce} benchmarking applications. 
Each job was assigned two servers that are connected to distinct groups of leaf switches, 
ensuring that the traffic between workers traversed the spine switches. 
With 8 concurrent jobs in play, the network bandwidth was expected to be saturated by these parallel flows.
Under these conditions, any suboptimal path selection could lead to an imbalanced traffic distribution,
with some of the network links are overloaded.
Such imbalances can severely impact the performance of collective communication, 
making the role of C4P crucial in maintaining peak communication performance and preventing potential bottlenecks within the network infrastructure.

\begin{figure}[t]
    \centering
    \subfloat[Allreduce Performance in the network with 1:1 oversubscription.]{
        \centering
        \includegraphics[width=0.75\linewidth]{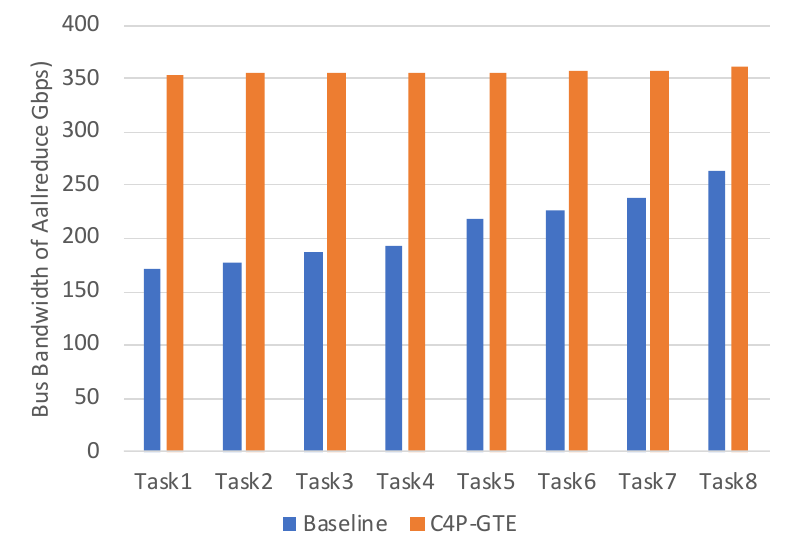}
        \label{fig:gtea}}
     \hfill
    \subfloat[Allreduce Performance in the network with 2:1 oversubscription.]{
        \centering
        \includegraphics[width=0.75\linewidth]{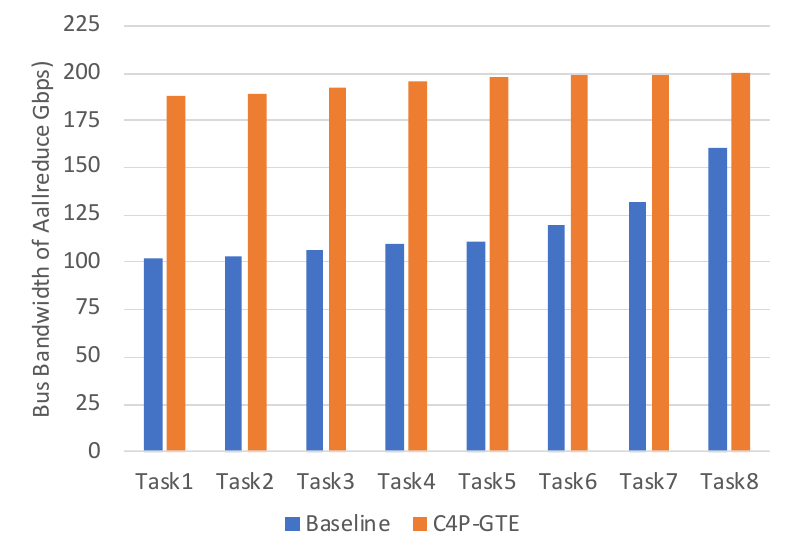}
        \label{fig:gteb}}
    \caption{Effectiveness of global traffic engineering.}
    \vspace{-0.5cm}
    \label{fig:gte}
\end{figure}

Fig.~\ref{fig:gtea} illustrates the performance outcomes of running concurrent \emph{allreduce} 
benchmarks within a network that maintains a 1:1 oversubscription rate. 
The data indicates that all tasks exhibit similar levels of performance with C4P enabling global traffic engineering, 
closely approaching the maximum achievable throughput for a single task on this testbed. 
Specifically, the performance metrics for these tasks range between 353.86 Gbps and 360.57 Gbps.
In contrast, when C4P is not enabled, there is a substantial performance degradation and variation among the tasks. 
The lowest-performing task achieves only 171.93 Gbps, while the best-performing task reaches 263.27 Gbps. 
On average, C4P improves the overall system throughput by 70.3\%.
This comparison clearly highlights the substantial improvements in bandwidth utilization and consistency enabled by C4P.

\begin{figure}[t]
\centering
\includegraphics[width=\linewidth]{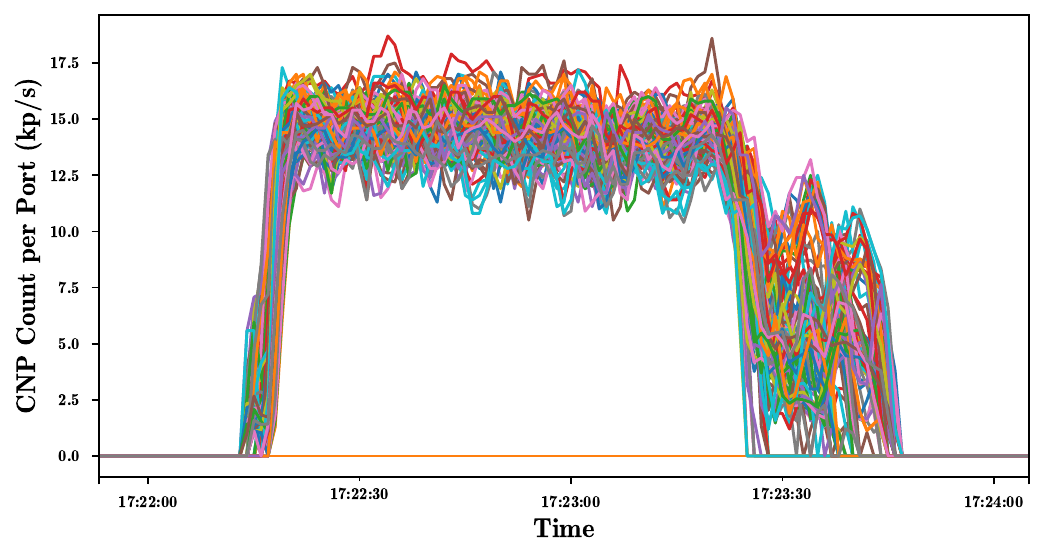}
\caption{The CNP count received in each bonded port.}
\vspace{-0.5cm}
\label{fig:cnpcount}
\end{figure}

It is important to clarify that the peak throughput in this system is constrained by the NVLink fabric, 
causing the maximum attainable bus bandwidth at 362 Gbps. 
Because of the limitation of NVLink fabric, the network's capacity is underutilized, 
which results in an absence of queue buildup in switches and congestion-induced transmission rate reduction in senders.
To evaluate the performance of C4P within a congested network environment, 
we intentionally reduced the number of active spine switches by half, 
effectively increasing the network oversubscription rate to 2:1. 
Under this new configuration, we replicated the previous experiments to observe how C4P handles increased traffic conditions. 
The outcomes of these tests are presented in Fig.~\ref{fig:gteb}.
A similar conclusion can be drawn from these outcomes, 
with the exception that the performance of concurrent tasks exhibits slight variations, when C4P is enabled.
Specifically, we observe a small gap of just 11.27 Gbps between the highest and lowest throughputs.
This difference is attributed to the congestion control mechanism in RDMA NICs, 
which dynamically adjusts the data transmission rate at the sender side in response to network conditions.
As depicted in Fig.~\ref{fig:cnpcount}, each bonded port receives approximately 15,000 Congestion Notification Packets (CNPs) every second,
with the count fluctuating between 12,500 and 17,500.
The arrival of CNPs instigates a throttling of the data transfer rate from the senders.
Fluctuations in CNP receiving frequency directly contribute to the observed variability in the flows' effective bandwidth.
Despite these variations, the overall throughput is substantially improved by 65.55\%, and the long-tail problem is obviously alleviated.

\begin{figure}[t]
    \centering
    \subfloat[Allreduce Performance with C4P static traffic engineer.]{
        \centering
        \includegraphics[width=0.9\linewidth]{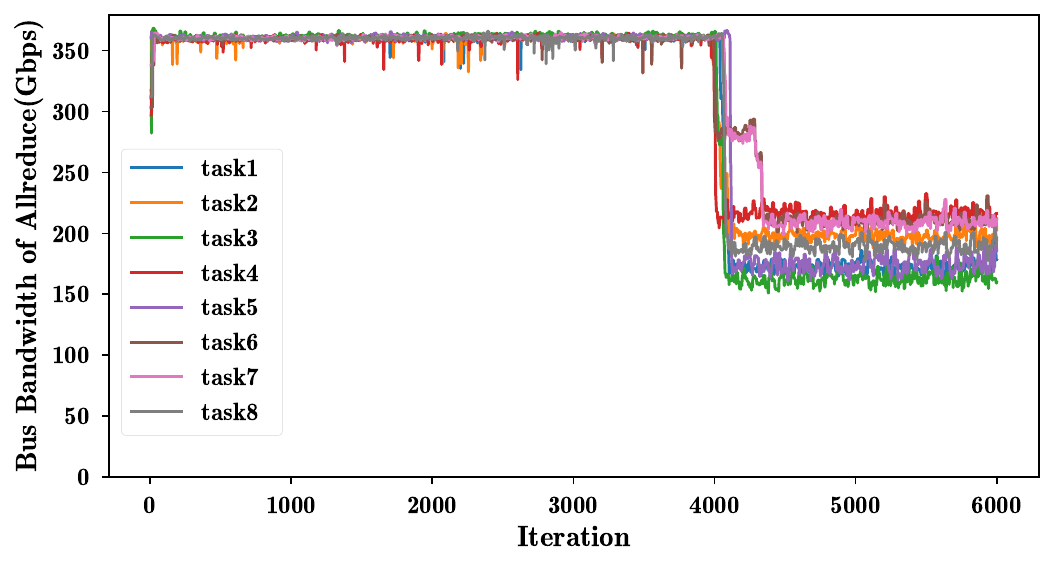}
        \label{fig:archa}}
     \hfill
    \subfloat[Allreduce Performance with C4P dynamic load balance.]{
        \centering
        \includegraphics[width=0.9\linewidth]{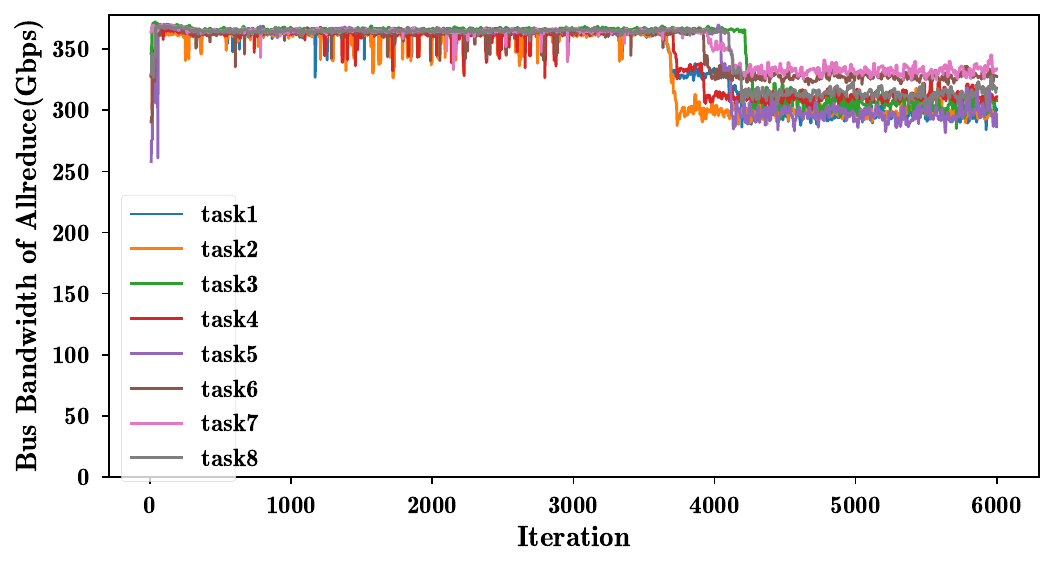}
        \label{fig:archb}}
    \caption{Effectiveness of load balance when link failure appears.}
    \vspace{-0.5cm}
    \label{fig:linkdown}
\end{figure}

\textbf{Tolerance to dynamic link failures.}
The efficacy of global traffic engineering relies on the stability of the underlying network conditions. 
A down-link can necessitate the rerouting of impacted flows, which could lead to imbalanced traffic distribution within the network,
and dismiss the efficacy of global traffic engineering.
In this case, load balance mechanism would take into effect to adjust the load on each flow, thereby rebalancing network traffic.
To assess the effectiveness of this mechanism, we replicated previous experiments in a 1:1 oversubscribed network and intentionally deactivated a link during the experiments. 
To obtain the tasks' instant performance, other than averaged value after the tasks are finished, 
we refined our testing benchmark to output results immediately upon completion of each \emph{allreduce} operation.
And the results are shown in Fig.~\ref{fig:linkdown}.
Though high-frequency timestamp recording results in some fluctuations in the data, 
we can still observe significant improvement in system throughput, with dynamic load balance enabled.
In the tests that C4P load balancing is not activated, 
the bus bandwidth for the given tasks experiences considerable degradation, 
with values ranging between 160 Gbps and 220 Gbps and an average bandwidth of 185.76 Gbps. 
Conversely, when C4P load balancing is enabled, the observed bus bandwidth improves significantly, 
ranging from 290 Gbps to 335 Gbps, with an average of 301.46 Gbps. 
This enhancement represents a substantial performance gain of 62.3\% when a link error occurs during the training processes.
Note that, with 1 link error among the 8 uplinks, the theoretical ideal performance would be 7/8 of the original, i.e., 315 Gbps. 
And the performance achieved with C4P enabled is found to be in close proximity to this ideal value.

\begin{figure}[t]
    \centering
    \subfloat[Bandwidth per switch port with C4P static traffic engineer.]{
        \centering
        \includegraphics[width=0.9\linewidth]{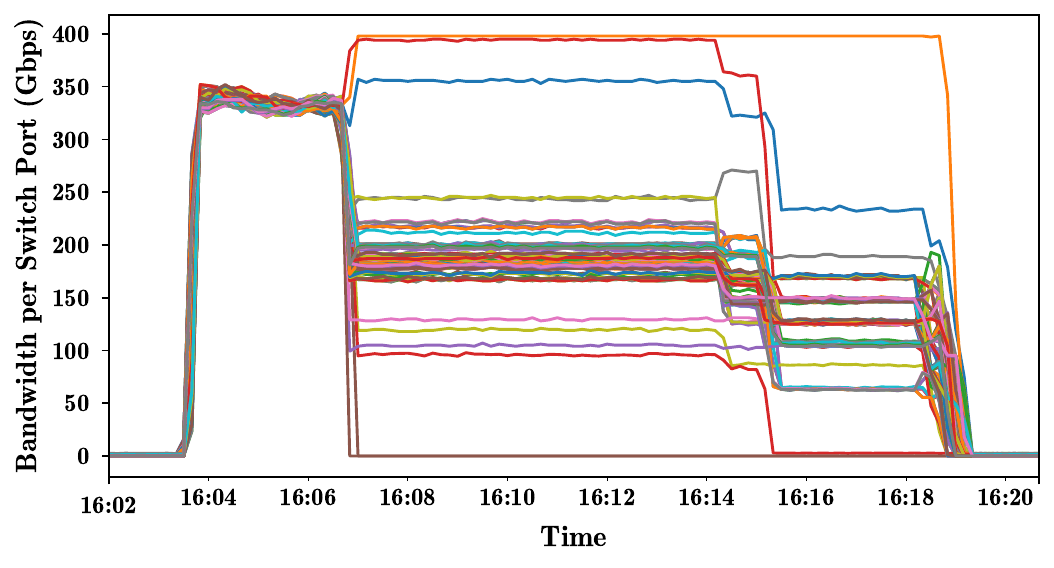}
        \label{fig:bwa}}
     \hfill
    \subfloat[Bandwidth per switch port  with C4P dynamic load balance.]{
        \centering
        \includegraphics[width=0.9\linewidth]{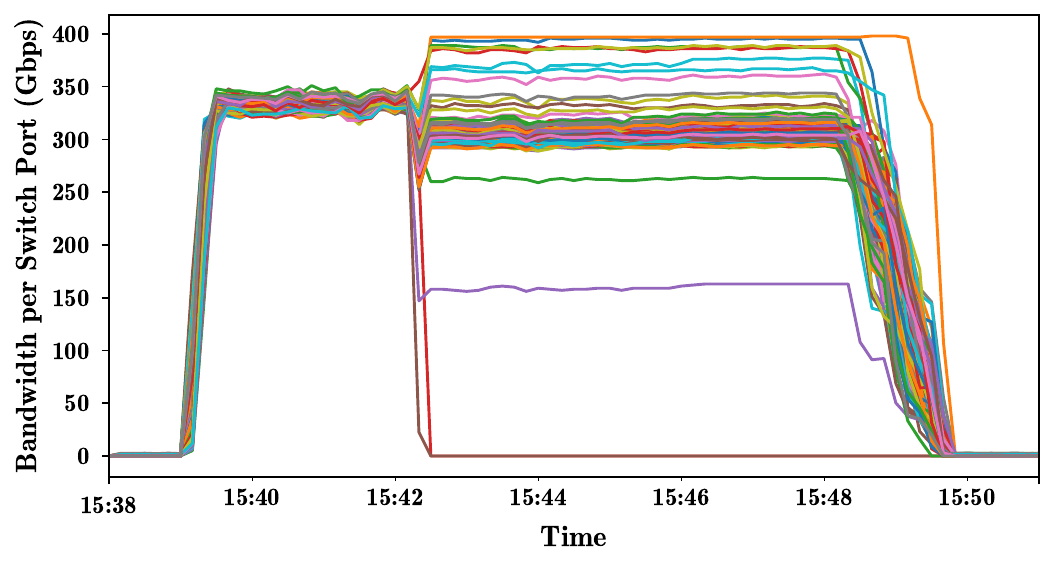}
        \label{fig:bwb}}
    \caption{Comparison on switch port bandwidth with/without load balance.}
    \label{fig:portbw}
\end{figure}

To demonstrate how C4P load balance improves the system throughput, 
we have collected the statistics from the leaf switches regarding the traffic distribution across each port, as illustrated in Fig.~\ref{fig:portbw}.
With C4P traffic engineering implemented, all upstream ports exhibit near-optimal bandwidth utilization prior to the induction of a link error.
However, the emergence of the link error leads to a marked discrepancy in throughput across different ports.
In the absence of C4P load balancing, only three of the ports exhibit traffic increment, 
indicating that the flows intended for the disabled link are being rerouted to these ports.
As a result, the available bandwidth for each flow in these congested ports is reduced, 
and the performance of the flows in the same communication channel is also negatively affected.
Consequently, a noticeable reduction in bandwidth is observed in the other ports.
Conversely, C4P load balancing dynamically adjusts the distribution of network traffic, 
decreasing the burden on congested flows while increasing it on those underutilized paths. 
This strategy leads to a more balanced load across the healthy links, and results in an overall enhancement of the system's throughput.

\textbf{Performance improvement in real-life jobs.}
To evaluate the efficacy of C4P in enhancing the performance of real-world applications, we conducted tests using three representative training jobs:
Job1 involves training a GPT model utilizing the Megatron framework, incorporating both TP and DP strategies for distributed training. 
This model comprises 22 billion parameters, and the TP and DP sizes are set to 8 and 16, respectively.
Job2 trains a Llama model, which contains 7 billion parameters. 
For this task, the Deepspeed framework is employed for distributed training. 
The zero-optimization~\cite{zero} is activated, with the training leveraging only DP.
Job3 entails training another GPT model with 175 billion parameters. 
This model is also trained based on Megatron framework, using both TP and PP strategies. 
With both TP and PP sizes configured as 8, it effectively creates 2 DP groups.

\begin{figure}[t]
\centering
\includegraphics[width=0.75\linewidth]{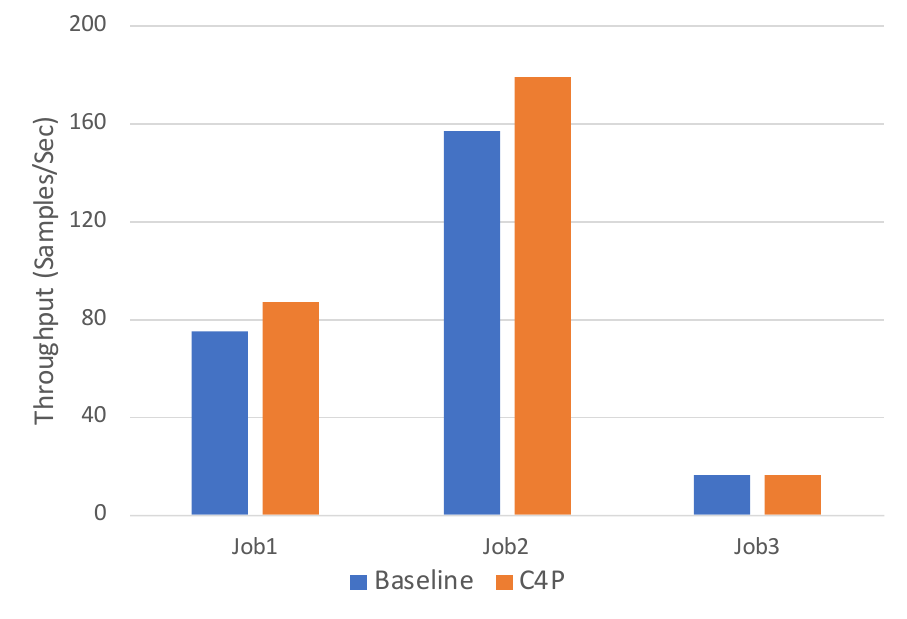}
\caption{Performance improvement in real-life jobs.}
\vspace{-0.5cm}
\label{fig:job}
\end{figure}

The performance evaluation results are depicted in Fig.~\ref{fig:job}. 
As can be discerned from the figure, significant improvements in the performance of the first two jobs. 
Specifically, the throughput of Job1 has increased by 15.95\%, rising from 74.82 to 86.76 samples per second. 
Similarly, Job2 shows an enhancement of 14.1\%, with throughput climbing from 156.59 to 178.65 samples per second. 
In contrast, Job3 does not exhibit a noticeable improvement in performance.
Our analysis suggests that the discrepancy in performance gains is closely associated with the proportion of communication time within a training step. 
For the first two jobs, the communication overhead constitutes over 30\% of each iteration's duration. 
However, Job3 is configured with a high Gradient Accumulation (GA) value of 16, 
which implies the communication cost is reduced by 16x. 
This setting substantially reduces the relative communication cost, which explains the minimal performance improvement observed for Job3.

\section{Related Work and Discussion}
To better utilize GPU resources in LLM training, existing research makes effort in two dimensions: stability improvement and performance improvement.

\textbf{Improving stability by monitoring and fault tolerance.}
Existing research on data center monitoring primarily focuses 
on general applications within data centers~\cite{yu2015software,joshi2018burstradar,kutare2010monalytics}, 
focusing on the behavior of hardware components 
such as CPUs, RDMA NICs, and switches. 
However, when compared to conventional data center applications, 
the training of LLMs is characteristically on a much larger scale 
and involves more complex interactions between software and hardware 
modules. This complexity necessitates more sophisticated, 
high-granularity monitoring solutions~\cite{rasley2020deepspeed,nsight} 
that are integrated within the software stack.

In terms of fault tolerance, 
the prevalent approach is the use of in-memory
checkpointing~\cite{wang2023gemini,nicolae2020deepfreeze,mohan2021checkfreq,eisenman2022check}. 
These efforts copy the current LLM states to the host memory with high
frequency such as one checkpoint per iteration.
In the event of an error, once the
issue has been identified, 
training can resume from the most
recent checkpoint. 

In addressing the challenge of straggler components, Jiang et al.~\cite{megascale} proposed an optimization solution to enhance the efficiency of large-scale AI clusters. 
Specifically, they identified slow components by monitoring the execution time of specific computing kernels. 
However, this approach is not suitable for cloud environments.

\textbf{Improving performance by better training architectures and network.}
With given computation resources, 
the performance of LLM training can be
improved by multiple-layer efforts, 
including (from the higher layer to the
lower layer) training architectures, 
communication libraries, and network structure. 
The LLM training tries to parallelize GPU computation 
with communication and CPU operations~\cite{zhang2023cocktailer,rasley2020deepspeed,shoeybi2019megatron}, 
or change the behavior of LLM training to reduce the overload, 
leveraging the properties of specific model structures~\cite{cui2023optimizing,liu2023janus}.
Communication libraries 
try to better utilize the given network bandwidth, 
including NVLink, PCIe, and RNIC, to shorten the delay of a data transmission~\cite{jeaugey2017nccl,msccl,accl,gloo}, or design better strategies to aggregate data in collective communication~\cite{shah2023taccl,cowan2023mscclang,cai2021synthesizing}.
Network structure research explore reasonable topologies to connect multiple hosts with appropriate bandwidth, providing better performance in LLM training.


Adaptive routing~\cite{dally-routing} and packet spraying~\cite{packet-spray} generally point to the same solution, which involves performing per-packet load balancing within switches based on network conditions. 
The decision to implement adaptive routing is entirely up to the end users, who must configure the related condition variables of CCL.
However, this approach poses significant challenges, as the efficiency of adaptive routing can be compromised by the flows that are deterministic routed.

\textbf{Applicability and Limitations of C4.}
C4D has broad applicability in business AI training tasks that rely on collective communication. Although we demonstrate its effectiveness using DP as an example, it is natural to wonder whether it can also be used in other scenarios, such as Pipeline Parallel (PP)~\cite{pp} and Expert Parallel (EP)~\cite{ep}.  
While C4D may not be able to detect non-communication hang or slow events within PP stages, as it relies on send/recv operations rather than collective communication, any resulting hang or slow effects would propagate to subsequent DP stages and could still be caught by C4D. 
In the case of EP, load imbalance among workers may occur, which can be mitigated by averaging collected data over a predefined period to smooth out random variations and highlight systemic issues. In future work, we plan to incorporate load variation into C4D to enable more precise slow detection.

The primary limitations of C4 includes: C4D’s error detection capabilities rely on syndromes observed through our enhanced communication library. Consequently, C4D has limited ability to detect errors occurring during the job initialization phase, before collective communication operations are initiated.
And deploying C4P requires detailed knowledge of the system configuration, including network topology, the number of switches at each layer, the number of ports per switch. Adapting C4P to different clusters necessitates retrieving this information from a background management system.

\section{Acknowledgment}
This work was partially supported by Research Grants Council of HKSAR (16213824). We thank Yu Zhou for developing the initial version of C4D.

\section{Conclusion}
Achieving optimal performance in training LLMs within large-scale AI clusters is challenging, hindered by frequent errors and traffic congestion.
To address these issues, we have developed C4, a communication-centric solution designed to minimize system downtime and enhance communication efficiency. 
By leveraging the characteristics of distributed training, \sys can substantially reduce the system recovery cost when faced with non-correctable errors.
It achieves this through rapid online fault detection, immediate isolation of anomalies, and by working in synergy with regular checkpoints and automated restart mechanisms.
Additionally, C4 employs the predictable patterns of collective communication to implement precise traffic management, which greatly alleviates network congestion.
\sys can reduce error-induced overhead by roughly 30\% and improve their throughput by about 15\% for tested applications with moderate communication costs.


\bibliographystyle{IEEEtranS}
\bibliography{ref}

\end{document}